\documentclass[journal]{IEEEtran}
\ifCLASSINFOpdf
  \usepackage[pdftex]{graphicx}
\else
\fi
\usepackage{cite}
\usepackage{amsmath,amssymb,amsfonts}
\usepackage{algorithmic}
\usepackage{graphicx}
\usepackage{textcomp}
\usepackage{xcolor}
\usepackage{amsmath}
\usepackage{algorithmic}
\usepackage{array}
\usepackage{stfloats}
\begin{document}
\title{IQ Skew and Imbalance Estimation for Coherent Point-to-Multi-Point Optical Networks}

\author{Ji Zhou, Jianrui Zeng, Haide Wang, Dong Guo, Liangchuan Li, Weiping Liu, and  Changyuan Yu
\thanks{Manuscript received; revised. This work was supported in part by the National Key R\&D Program of China under Grant 2023YFB2905700, in part by the National Natural Science Foundation of China under Grant 62371207 and Grant 62005102, and in part by the Young Elite Scientists Sponsorship Program by CAST under Grant 2023QNRC001. \it{(Corresponding author:Dong Guo and Liangchuan Li.)}}
\thanks{Ji Zhou, Jianrui Zeng, Haide Wang, and Weiping Liu are with Department of Electronic Engineering, College of Information Science and Technology, Jinan University, Guangzhou 510632, China.}
\thanks{Dong Guo is with School of Information and Electronics, Beijing Institute of Technology, Beijing 100081, China.}
\thanks{Liangchuan Li are with Optical Research Department, Huawei Technologies Co Ltd, Dongguan, 523808, China.}
\thanks{Changyuan Yu is with Department of Electrical and Electronic Engineering, The Hong Kong Polytechnic University, Hong Kong.}}

\maketitle
\begin{abstract} \boldmath
Coherent point-to-multi-point (PtMP) optical network based on digital subcarrier multiplexing (DSCM) has been a promising technology for metro and access networks to achieve cost savings, low latency, and high flexibility. In-phase and quadrature (IQ) impairments of the coherent transceiver (e.g. IQ skew and power imbalance) cause severe performance degradation. In the DSCM-based coherent PtMP optical networks, it is hard to realize far-end IQ-impairments estimation for the hub transmitter because the leaf on one subcarrier cannot acquire the signal on the symmetrical subcarrier. In this paper, we propose a far-end IQ-impairments estimation based on the specially designed time-and-frequency interleaving tones (TFITs), which can simultaneously estimate IQ skews and power imbalances of the hub transmitter and leaf receiver at an individual leaf. The feasibility of the TFITs-based IQ-impairments estimation has been experimentally verified by setting up $8$Gbaud/SC×$4$SCs DSCM-based coherent PtMP optical network. The experimental results depict that the absolute errors in the estimated IQ skew and power imbalance are within $\pm 0.5$ps and $\pm 0.2$dB, respectively. In conclusion, TFITs-based IQ-impairments estimation has great potential for DSCM-based coherent PtMP optical networks.
\end{abstract}

\IEEEpeerreviewmaketitle

\section{Introduction}
\IEEEPARstart{D}RIVEN by emerging network services, hub and spoke (H\text{\&}S) networks have become the main network architecture for access and metro networks \cite{campos2023coherent, zhang2023planning, castro2023scalable}. Coherent point-to-multi-point (PtMP) optical networks enabled by digital subcarrier multiplexing (DSCM) are the optimal solutions for matching the requirements of H\text{\&}S networks \cite{welch2021point, welch2023digital, rashidinejad2023real}. Compared to coherent point-to-point (P2P) optical networks, the DSCM-based coherent PtMP optical networks can provide massive connections between a high-speed hub node and multiple low-speed leaf nodes, achieving cost savings, lower latency, and higher flexibility \cite{pavon2022benefits, pavon2023tree, xing2023first}. However, in-phase and quadrature (IQ) impairments of the coherent transceiver (e.g. IQ skew and power imbalance) cause image crosstalk between the symmetrical subcarriers of the DSCM signal. Particularly, the image crosstalk caused by IQ skew grows with the increase of the subcarrier frequency \cite{zhang2019800g, duthel2023dsp, tong2023experimental}. Therefore, the image crosstalk causes significant performance degradation on the leaves carried by the high-frequency subcarriers \cite{bosco2016impact, fan2023hardware, duthel2023dsp-1}.

The IQ impairments of the coherent transceiver have been widely studied for coherent P2P optical networks. The transmitter (Tx) IQ impairments can be extracted from the tap coefficients of the multiple-input multiple-output (MIMO) equalizer, which has been widely applied in the single-carrier coherent P2P optical networks \cite{fan2018experimental, liang2019transceiver, fan2019transceiver}. Furthermore, the 4$\times$4 MIMO equalizer has been recently proposed to estimate Tx IQ impairments for DSCM-based coherent P2P optical networks by processing symmetrical subcarrier pairs \cite{zhai2020transmitter, ng2022far}. In addition, Tx IQ-impairments estimation based on a specially designed training sequence is proposed for coherent P2P optical networks. The uniformly-spaced frequency tones are used to acquire the Tx IQ skew by calculating the slope of their phase difference \cite{dai2022experimental, dai2022simultaneously}. Another method based on the single sideband comb signals is proposed to estimate the Tx IQ impairments by analyzing the image spectrum\cite{chen2017accurate}.

\begin{figure}[!t]
\centering
\includegraphics[width=\linewidth]{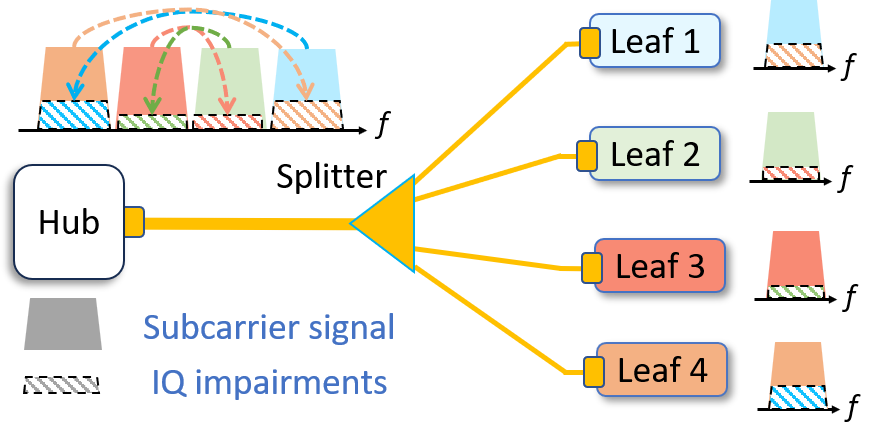}
\caption{Schematic diagram of DSCM-based coherent PtMP optical networks with Tx IQ impairments.}
\label{fig_1}
\end{figure}

Fig. \ref{fig_1} shows the schematic diagram of coherent PtMP optical networks based on DSCM with Tx IQ impairments. The hub transmits the DSCM signal with four subcarriers to the leaves, and one leaf receives its corresponding subcarrier rather than symmetrical subcarrier pairs. The above-mentioned schemes for coherent P2P optical networks require the whole spectrum of the transmitted signal to estimate the Tx IQ impairments, which are difficult to apply in DSCM-based coherent PtMP optical networks. This paper proposes the first far-end IQ-impairments estimation for the DSCM-based coherent PtMP optical networks, which can accurately estimate IQ skews and power imbalances of the hub Tx and leaf receiver (Rx) at the leaf. The main contributions of the work are as follows:
\begin{itemize}
\item Time-and-frequency interleaving tones (TFITs) are specially designed to accurately estimate the IQ impairments for the DSCM-based coherent PtMP optical networks, which solves the challenge of the far-end IQ-impairments estimation at an individual leaf.
\item $8$Gbaud/SC×$4$SCs DSCM-based coherent PtMP optical network is experimentally demonstrated to verify the feasibility of the TFITs-based IQ-impairments estimation. The absolute errors in the estimated IQ skew and power imbalance are within $\pm 0.5$ps and $\pm 0.2$dB, respectively.
\end{itemize}

\begin{figure}[!t]
\centering
\includegraphics[width=\linewidth]{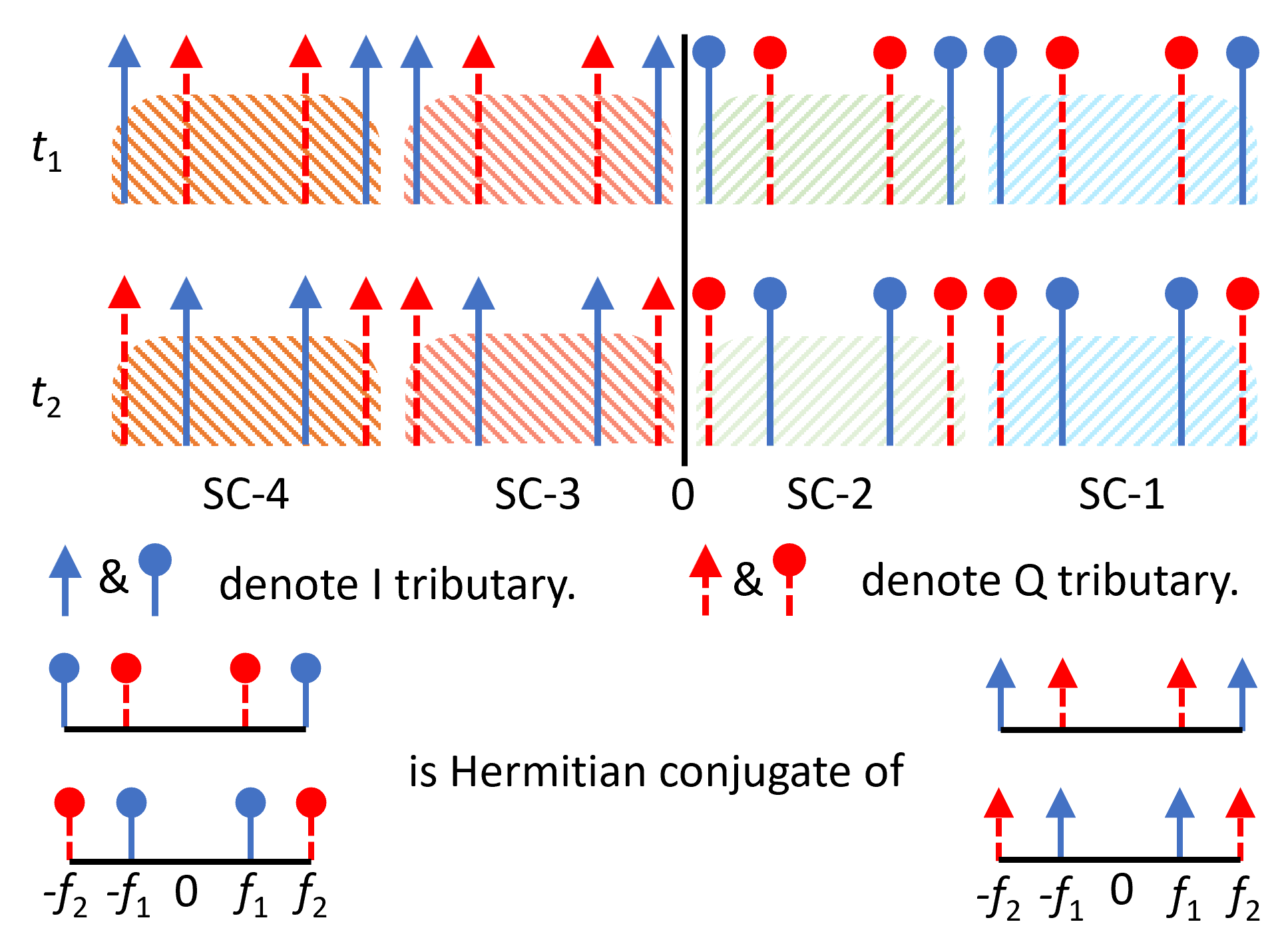}
\caption{Structure of the specially designed TFITs for one polarization.}
\label{fig_2}
\end{figure}

The remainder of the paper is organized as follows. In Section \text{II}, the TFITs structure is designed to estimate IQ impairments. The principle of IQ-impairments estimation is introduced for DSCM-based coherent PtMP optical networks. In Section \text{III}, we introduce the experimental setups of a DSCM-based coherent PtMP optical network. In Section \text{IV}, the experimental results and discussions are demonstrated to verify the performance of the TFITs-based IQ-impairments estimation. Finally, the paper is concluded in Section \text{V}. 

\section{TFITs Structure and Estimation Principle}
In this section, the TFITs structure for IQ-impairments estimation is designed for the DSCM-based coherent PtMP optical networks. Then the TFITs affected by IQ impairments are analyzed. Finally, the principle of TFITs-based far-end IQ-impairments estimation is given.

\begin{figure}[!t]
\centering
\includegraphics[width=0.96\linewidth]{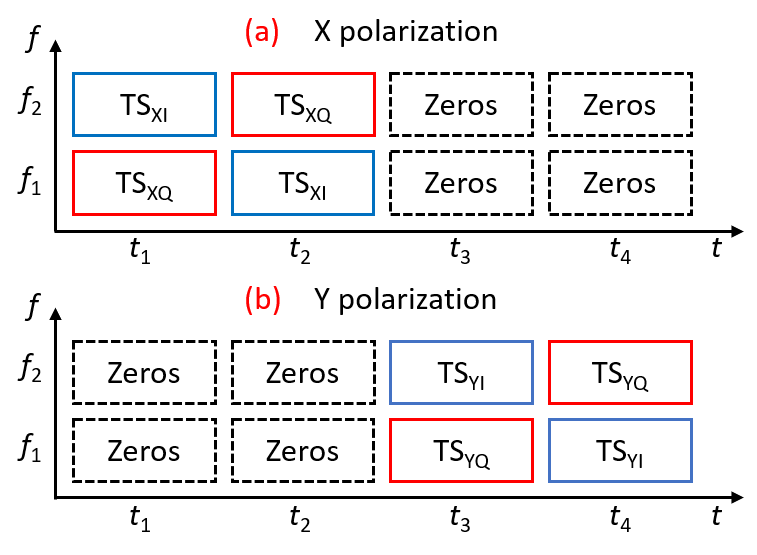}
\caption{The specially designed TFITs including four time slots $t_1$, $t_2$, $t_3$, $t_4$ and two tone frequencies $f_1$, $f_2$ for (a) X polarization and (b) Y polarization.}
\label{fig_3}
\end{figure}

\subsection{TFITs Design for IQ-impairments estimation}
The specially designed TFITs structure for one polarization is shown in Fig. \ref{fig_2}. The first subcarrier (SC-1) and fourth subcarrier (SC-4) are a pair of symmetrical subcarriers, while the second subcarrier (SC-2) and the third subcarrier (SC-3) are a pair of symmetrical subcarriers. Taking the symmetrical SC-1 and SC-4 into consideration, after the frequency shift, the intermediate frequency signal $s_1(t)$ of SC-1 and $s_4(t)$ of SC-4 can be expressed as
\begin{equation}\label{eq1}
    s_1(t)=[I_1(t)+jQ_1(t)]e^{+j2\pi f_\text{c} t}
\end{equation}
and
\begin{equation}\label{eq2}
    s_4(t)=[I_4(t)+j Q_4(t)]e^{-j2\pi f_\text{c} t},
\end{equation}
respectively. The center frequencies of the SC-1 and SC-4 are $+f_\text{c}$ and $-f_\text{c}$, respectively. $I_1(t)+jQ_1(t)$ and $I_4(t)+j Q_4(t)$ are the baseband signal on the SC-1 and SC-4, respectively. Therefore, the training sequence (TS) $s_{\text{Tx}}(t)$ is generated as
\begin{equation}\label{eq3}
s_{\text{Tx}}(t)=s_1(t)+s_4(t)=s_{\text{Tx},\text{I}}(t)+js_{\text{Tx},\text{Q}}(t)
\end{equation}
where $s_{\text{Tx},\text{I}}(t)$ and $s_{\text{Tx},\text{Q}}(t)$ are the I and Q tributaries of the TS $s_{\text{Tx}}(t)$, which can be expressed as
\begin{equation}\label{eq4}
s_{\text{Tx},\text{I}}(t)=[I_1(t)+I_4(t)]\cos2\pi f_\text{c} t+[Q_4(t)-Q_1(t)]\sin 2\pi f_\text{c} t
\end{equation}
and 
\begin{equation}\label{eq5}
s_{\text{Tx},\text{Q}}(t)=[I_1(t)-I_4(t)]\sin 2\pi f_\text{c} t+[Q_1(t)+Q_4(t)]\cos 2\pi f_\text{c} t,
\end{equation}
respectively. As Fig. \ref{fig_2} shows, each subcarrier consists of two tones at the frequencies $f_1$ and $f_2$, which can be defined as $I_{f_1}(t)+jQ_{f_1}(t)$ and $I_{f_2}(t)+jQ_{f_2}(t)$, respectively. in the time slot $t_1$, the tone at the frequency $f_1$ inserted on SC-4 is the Hermitian conjugate of that inserted on SC-1. In other word, $I_1(t)+jQ_1(t)$ is equal to $-I_4(t)+jQ_4(t)$ where $I_1(t)$ and $Q_1(t)$ are equal to $I_{f_1}(t)$ and $Q_{f_1}(t)$, respectively. Therefore, the Q tributary of TFITs with the frequency $f_1 + f_\text{c}$ tone can be expressed as 
\begin{equation}\label{eq6}
    s_{\text{Tx,Q}}(t)=2\times \left[ I_{f_1}(t) \sin 2\pi f_\text{c} t+Q_{f_1}(t)\cos 2\pi f_\text{c} t \right].
\end{equation}
The I tributary of the TFITs with the frequency $f_1 + f_\text{c}$ tone is zero. Furthermore, the tone at frequency $f_2$ inserted on SC-4 is the Hermitian conjugate of that inserted on SC-1. In other word, $I_1(t)+jQ_1(t)$ is equal to $I_4(t)-jQ_4(t)$ where $I_1(t)$ and $Q_1(t)$ are equal to $I_{f_2}(t)$ and $Q_{f_2}(t)$, respectively. Therefore, the I tributary of TFITs with the frequency $f_2 +f_\text{c}$ tone can be expressed as 
\begin{equation} \label{eq7}
    s_{\text{Tx,I}}(t)=2\times \left[ I_{f_2}(t) \cos 2\pi f_\text{c} t-Q_{f_2}(t)\sin 2\pi f_\text{c} t \right].
\end{equation}
The Q tributary of the TFITs at frequency $f_2 + f_\text{c}$ is zero. 

The $f_1$ and $f_2$ can be set to any frequencies within $f_s$ where $f_s$ denotes the baud rate. Specifically, $f_1$ and $f_2$ are set to a quarter of the baud rate (i.e. $f_s/4$) and half of the baud rate (i.e. $f_s/2$), respectively. Thus, by combining the Eq. (\ref{eq6}) and Eq. (\ref{eq7}), the TFITs in the time slot $t_1$ can be expressed as
\begin{equation}\label{eq8}
s_{\text{Tx,}t_1}(t)=s_{\text{Tx,I}}(t)+js_{\text{Tx,Q}}(t) = s_{\text{Tx},{f_s/2}}(t)+js_{\text{Tx},{f_s/4}}(t).
\end{equation}
To eliminate the influence on IQ-impairments estimation caused by the phase-frequency response caused by the devices and chromatic dispersion, TFITs are also designed using two time slots with two interleaving frequencies. Compared to time slot $t_1$, the frequencies of the I and Q tributaries are interchanged in the time slot $t_2$. Therefore, the TFITs in the time slot $t_2$ can be expressed as
\begin{equation}\label{eq9}
s_{\text{Tx,}t_2}(t)=s_{\text{Tx,I}}(t)+js_{\text{Tx,Q}}(t) = s_{\text{Tx},{f_s/4}}(t)+js_{\text{Tx},{f_s/2}}(t).
\end{equation}
The TFITs can be generated by combining the signals of Eqs. (\ref{eq8}) and (\ref{eq9}). Fig. \ref{fig_3} shows the specially designed TFITs include four time slots [$t_1$, $t_2$, $t_3$, $t_4$] and two tones at frequencies [$f_1$, $f_2$] for (a) X and (b) Y polarizations. At the time slots of $t_1$ and $t_2$, the TFITs are carried on the X polarization, while the Y polarization is zero to avoid inter-polarization crosstalk between X and Y polarizations \cite{zhu2018reception, zhang2021efficient, liu2008intra}. At the time slots of $t_3$ and $t_4$, the proposed TFITs are carried on the Y polarization, while the X polarization is zero. 

\newcounter{TempEqCnt1}
\setcounter{TempEqCnt1}{\value{equation}}
\setcounter{equation}{12}
\begin{figure*}
\begin{equation}\label{eq13}
\begin{aligned}
s_{\text{Rx},f_s/2}(t)=& 2 \left[ I_{f_s/2}(t) \cos 2\pi f_\text{c} t-Q_{f_s/2}(t)\sin 2\pi f_\text{c} t \right]\times e^{-j \left[ 2 \pi (f_\text{c} +\Delta f)t+\phi\right]}\\
=& \underbrace{\begin{bmatrix} I_{f_s/2}(t)& -Q_{f_s/2}(t) \end{bmatrix}\begin{bmatrix} \cos\phi_1(t)\\ \sin \phi_1(t)\end{bmatrix}}_{ \text{$r_{f_s/2, \text{I}}(t)$}}+
j\underbrace{\begin{bmatrix} I_{f_s/2}(t)& Q_{f_s/2}(t) \end{bmatrix}\begin{bmatrix} \sin\phi_1(t)\\ \cos \phi_1(t)\end{bmatrix}}_{ \text{$r_{f_s/2, \text{Q}}(t)$}}
\end{aligned}
\end{equation}
\hrulefill
\end{figure*}
\setcounter{equation}{\value{TempEqCnt1}}
\newcounter{TempEqCnt2}
\setcounter{TempEqCnt2}{\value{equation}}
\setcounter{equation}{13}
\begin{figure*}
\begin{equation}\label{eq14}
\begin{aligned}
js_{\text{Rx},f_s/4}(t)
= & 2jg_{\text{Tx}}\left[ I_{f_s/4}(t+\tau_{\text{Tx}}) \sin 2\pi f_\text{c} (t+\tau_{\text{Tx}})+ Q_{f_s/4}(t+\tau_{\text{Tx}})\cos 2\pi f_\text{c} (t+\tau_{\text{Tx}}) \right] e^{-j \left[ 2 \pi (f_\text{c} +\Delta f)t+\phi\right]}\\
= & \underbrace{g_{\text{Tx}}\begin{bmatrix} I_{f_s/4}(t+\tau_{\text{Tx}})& -Q_{f_s/4}(t+\tau_{\text{Tx}}) \end{bmatrix}\begin{bmatrix} \cos\phi_2(t)\\ \sin \phi_2(t)\end{bmatrix}}_{\text{$r_{f_s/4, \text{I}}(t)$}}+
j\underbrace{g_{\text{Tx}}\begin{bmatrix} I_{f_s/4}(t+\tau_{\text{Tx}})& Q_{f_s/4}(t+\tau_{\text{Tx}}) \end{bmatrix}\begin{bmatrix} \sin\phi_2(t)\\ \cos \phi_2(t)\end{bmatrix}}_{\text{$r_{f_s/4, \text{Q}}(t)$}}
\end{aligned}
\end{equation}
\hrulefill
\end{figure*}
\setcounter{equation}{\value{TempEqCnt2}}

\subsection{The Model of IQ Impairments on TFITs}
Without loss of generality, we investigate the model of IQ impairment on the TFITs in the time slot $t_1$. The TFITs with Tx IQ impairments can be modeled as \cite{zhang2020multi}
\begin{equation}\label{eq10}
s_{\text{Tx}}(t)=s_{\text{Tx},{f_s/2}}(t)+jg_{\text{Tx}}s_{\text{Tx},{f_s/4}}(t+\tau_{\text{Tx}})
\end{equation}
where $\tau_{\text{Tx}}$ and $g^2_{\text{Tx}}$ represent the Tx IQ skew and power imbalance, respectively. After optical modulation, the launch optical signal can be defined as
\begin{equation}\label{eq11}
s_{\text{Tx, O}}(t)=\left[s_{\text{Tx},{f_s/2}}(t)+js'_{\text{Tx},{f_s/4}}(t)\right]e^{j2\pi f_0t}
\end{equation}
where $f_0$ is the center frequency of the optical carrier. $s'_{\text{Tx},f_s/4}(t)$ is defined as $g_{\text{Tx}}s_{\text{Tx},f_s/4}(t+\tau_{\text{Tx}})$.

At the Rx side, SC-1 or SC-4 can be selected by tuning the wavelength of the local oscillator (LO) laser to the frequencies of $f_0+f_\text{c}$ or $f_0-f_\text{c}$, respectively. For SC-1, $s_{\text{Tx, O}}(t)$ is multiplied by $e^{-j \left[ 2 \pi (f_0+f_\text{c} +\Delta f)t+\phi\right]}$ mathematically, where $\Delta f$ and $\phi$ are the frequency offset and phase noise of the laser, respectively. After frequency downshift, the received signal of SC-1 is expressed as
\begin{equation}\label{eq12}
\begin{aligned}
s_{\text{Rx}}(t)=&\left[s_{\text{Tx},{f_s/2}}(t)+js'_{\text{Tx},f_s/4}(t)\right]e^{-j \left[ 2 \pi (f_\text{c} +\Delta f)t+\phi\right]}\\=& s_{\text{Rx},f_s/2}(t)+js_{\text{Rx},f_s/4}(t)
\end{aligned}
\end{equation}
where $s_{\text{Rx},f_s/2}(t)$ and $js_{\text{Rx},f_s/4}(t)$ are denoted as Eqs. (\ref{eq13}) and (\ref{eq14}) show at the top of this page, respectively. $\phi_1(t)$ is $-(2 \pi \Delta f t + \phi)$ and $\phi_2(t)$ is $-(2 \pi \Delta f t + \phi-2 \pi f_\text{c} \tau_{\text{Tx}})$. After frequency downshift, both the I and Q tributaries of the received signal $s_{\text{Rx}}(t)$ contain the $f_s/4$ and $f_s/2$ tones, which can be expressed as 
\setcounter{equation}{14}
\begin{equation}\label{eq15}
s_{\text{Rx}}(t)= r_{f_s/2,\text{I}}(t)+r_{f_s/4,\text{I}}(t)+j\left[r_{f_s/2,\text{Q}}(t)+r_{f_s/4,\text{Q}}(t)\right]
\end{equation}
where $r_{f_s/2, \text{I}}(t)$, $r_{f_s/4, \text{I}}(t)$, $r_{f_s/2, \text{Q}}(t)$ and $r_{f_s/4, \text{Q}}(t)$ are denoted as Eqs. (\ref{eq13}) and (\ref{eq14}) show. Therefore, the I and Q tributaries of the signal $s_{\text{Rx}}(t)$ can be expressed as
\begin{equation}\label{eq16}
s_{\text{Rx,I}}(t)= r_{f_s/2, \text{I}}(t)+r_{f_s/4, \text{I}}(t)
\end{equation}
and
\begin{equation}\label{eq17}
s_{\text{Rx,Q}}(t)=r_{f_s/2, \text{Q}}(t)+r_{f_s/4, \text{Q}}(t),
\end{equation}
respectively. When the received signal suffers the Rx IQ impairments, it can be modeled as \cite{chung2010transmitter, wang2022correlation}
\begin{equation}\label{eq18}
s_{\text{Rx}}(t)= s_{\text{Rx,I}}(t)+jg_{\text{Rx}}s_{\text{Rx,Q}}(t+\tau_{\text{Rx}})
\end{equation}
where $\tau_{\text{Rx}}$ and $g_{\text{Rx}}^2$ denote the Rx IQ skew and power imbalance, respectively.

\begin{figure*}[t]
\centering
\includegraphics[width= \linewidth]{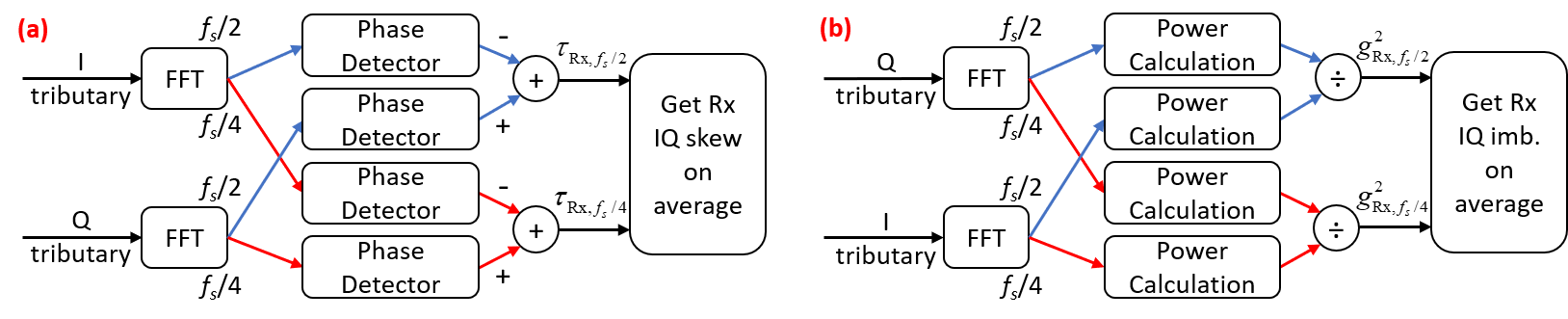}
\caption{Schematic diagram of IQ-impairments estimation based on the specially designed TFITs including Rx IQ (a) skew and (b) power imbalance.}
\label{fig_4}
\end{figure*}

\subsection{Estimation of Rx IQ Impairments in PtMP Networks}
According to the adding order of Tx and Rx IQ impairments, the Rx IQ impairments should be estimated and compensated before estimation for the Tx IQ impairments. Based on the model of Rx IQ impairments, we propose TFITs-based schemes for accurately estimating the Rx IQ skew and power imbalance.
\subsubsection{Estimation of Rx IQ Skew}
By substituting Eqs. (\ref{eq13}) and (\ref{eq14}) into Eq. (\ref{eq18}), it is inferred that in the time slot $t_1$, $r_{f_s/2, \text{I}}(t)$ contains only the timing phase offset caused by the out-of-synchronization clock, while $g_{\text{Rx}}r_{f_s/2, \text{Q}}(t+\tau_{\text{Rx}})$ contains both the timing phase offset of the clocks and Rx IQ skew $\tau_{\text{Rx}}$. Similarly, there is a difference of Rx IQ skew $\tau_{\text{Rx}}$ between the timing phase offsets of $r_{f_s/4, \text{I}}(t)$ and $g_{\text{Rx}}r_{f_s/4, \text{Q}}(t+\tau_{\text{Rx}})$. Therefore, the Rx IQ skew can be estimated by calculating the difference in the timing phase offset between the Q and the I tributaries of the tones with the same frequency. 


Fig. \ref{fig_4}(a) shows the schematic diagram of Rx IQ-skew estimation by using the I and Q tributaries of the Rx TFITs. By performing sampling and fast Fourier transform (FFT) on the I tributary of the received signal $s_{\text{Rx, I}}(t)$ to $S_\text{Rx, I}(k)$ and Q tributary $g_{\text{Rx}}s_{\text{Rx, Q}}(t+\tau_{\text{Rx}})$ to $S_\text{Rx, Q}(k)$ at each time slot, the timing phase offsets of I and Q tributaries can be calculated by the Godard phase detector using the $2n$ frequency points around $f_s/2$ \cite{godard1978passband}. Therefore, the Rx IQ skew is estimated by calculating the difference between timing phase offsets of the Q and I tributaries of $f_s/2$ tone, which can be expressed as 
\begin{equation}\label{eq19}
\begin{aligned}
\tau_{\text{Rx,}f_s/2}= & \frac{1}{2 \pi} \operatorname{arg} \left[\sum_{k=N / 4-n+1}^{N / 4+n} S_{\text{Rx, Q}}(k) S_{\text{Rx, Q}}^*\left(k+\frac{N}{2}\right)\right] \\
-&\frac{1}{2 \pi} \operatorname{arg} \left[\sum_{k=N / 4-n+1}^{N / 4+n} S_{\text{Rx, I}}(k) S_{\text{Rx, I}}^*\left(k+\frac{N}{2}\right)\right]
\end{aligned}
\end{equation}
where $\operatorname{arg}(\cdot)$ denotes phase extraction and $(\cdot)^*$ denotes conjugate operation. The FFT points of $S_{\text{Rx, I}}(k)$ and $S_{\text{Rx, Q}}(k)$ around the $N/4$ correspond to the frequency points around $f_s/2$. The timing phase offset can be also calculated by the modified Godard phase detector using the frequency points around $f_s/4$ \cite{wu2014frequency}. Therefore, the Rx IQ skew can be estimated by calculating the difference between timing phase offsets of the Q and I tributaries of $f_s/4$ tone, which is expressed as
\begin{equation}\label{eq20}
\begin{aligned}
\tau_{\text{Rx,}f_s/4}= & \frac{1}{\pi} \operatorname{arg} \left[\sum_{k=3N / 8-n+1}^{3N/8+n} S_{\text{Rx, Q}}(k) S_{\text{Rx, Q}}^*\left(k+\frac{N}{4}\right)\right] \\
-&\frac{1}{\pi} \operatorname{arg} \left[\sum_{k=3N/8-n+1}^{3N/8+n} S_{\text{Rx, I}}(k) S_{\text{Rx, I}}^*\left(k+\frac{N}{4}\right)\right]
\end{aligned}
\end{equation}
where the FFT points of $S_{\text{Rx, I}}(k)$ and $S_{\text{Rx, Q}}(k)$ around the $3N/8$ correspond to the frequency points around $f_s/4$. It is worth noting that the calculations for timing phase offset should be modified based on the frequencies $f_1$ and $f_2$.
Finally, the more accurate Rx IQ skew $\tau_{\text{Rx,IQ}}$ can be obtained by averaging $\tau_{\text{Rx,}f_s/2}$ and $\tau_{\text{Rx,}f_s/4}$.

\subsubsection{Estimation of Rx IQ Power Imbalance}
The Rx IQ power imbalance can be calculated by the power ratio between the Q and the I tributaries of the received signal, which can be defined as
\begin{equation}\label{eq21}
g_{\text{Rx,IQ}}^2=g_{\text{Rx}}^2 \frac{E\left[\left| s_{\text{Rx}, \text{Q}}\left(t+\tau_{\text{Rx}}\right)\right|^2\right]}{E\left[\left|s_{\text{Rx}, \text{I}}(t)\right|^2\right]}=g_{\text{Rx}}^2 \frac{E\left[\left| s_{\text{Rx}, \text{Q}}\left(t\right)\right|^2\right]}{E\left[\left|s_{\text{Rx}, \text{I}}(t)\right|^2\right]}.
\end{equation}
Due to the independence between the $f_s/2$ and $f_s/4$ tones, we can extract Rx IQ power imbalance using only the $f_s/2$ tone or $f_s/4$ tone. The power ratio between the Q and I tributaries of $f_s/2$ tone can be calculated by
\begin{equation}\label{eq22}
g_{\text{Rx,}f_s/2}^2=g_{\text{Rx}}^2 \frac{E\left[\left| r_{f_s/2, \text{Q}}\left(t\right)\right|^2\right]}{E\left[\left|r_{f_s/2, \text{I}}(t)\right|^2\right]}.
\end{equation}
The power ratio between the Q and I tributaries of $f_s/4$ tone can be calculated by
\begin{equation}\label{eq23}
g_{\text{Rx,}f_s/4}^2=g_{\text{Rx}}^2 \frac{E\left(\left|r_{f_s/4, \text{Q}}\left(t\right)\right|^2\right)}{E\left(\left|r_{f_s/4, \text{I}}(t)\right|^2\right)}.  
\end{equation}
The divisions of Eqs. (\ref{eq22}) and (\ref{eq23}) can be defined as
\begin{equation}\label{eq24}
\frac{E\left(\left| r_{f_s/2, \text{Q}}\left(t\right)\right|^2\right)}{E\left(\left|r_{f_s/2, \text{I}}(t)\right|^2\right)} = \frac{E\left(|I_{f_s/2}(t)|^2+|Q_{f_s/2}(t)|^2\right) + C_{f_s/2}}{E\left(|I_{f_s/2}(t)|^2+|Q_{f_s/2}(t)|^2\right)- C_{f_s/2}}
\end{equation}
and
\begin{equation}\label{eq25}
\frac{E\left(\left| r_{f_s/4, \text{Q}}\left(t\right)\right|^2\right)}{E\left(\left|r_{f_s/4, \text{I}}(t)\right|^2\right)} = \frac{E\left(|I_{f_s/4}(t)|^2+|Q_{f_s/4}(t)|^2\right)+ C_{f_s/4}}{E\left(|I_{f_s/4}(t)|^2+|Q_{f_s/4}(t)|^2\right)- C_{f_s/4}},
\end{equation}
respectively where
\begin{equation}\label{eq26}
 C_{f_s/2} = 2E\left[I_{f_s/2}(t) Q_{f_s/2}(t) \sin 2\phi_1(t) \right]
\end{equation}
and
\begin{equation}\label{eq27}
 C_{f_s/4} = 2E\left[I_{f_s/4}(t+\tau_{\text{Tx}}) Q_{f_s/4}(t+\tau_{\text{Tx}}) \sin 2\phi_2(t) \right].
\end{equation}
The existing frequency offset causes a varied $\phi_1(t)$ and $\phi_2(t)$ with a rotation of $2\pi$. Therefore, $ C_{f_s/2}$ and $C_{f_s/4}$ are equal to zero. In theory, the  $g_{\text{Rx,}f_s/2}^2$ and $g_{\text{Rx,}f_s/4}^2$ are both equal to the adding Rx IQ power imbalance $g_{\text{Rx}}^2$. 

Fig. \ref{fig_4}(b) shows the schematic diagram for Rx IQ power imbalance estimation by using the I and Q tributaries of the Rx TFITs. After FFT, two $f_s/2$ tones and two $f_s/4$ tones on the I and Q tributaries are extracted, respectively. According to Paswal's theorem, $g_{\text{Rx,}f_s/2}^2$ can be calculated by the power ratio between the two $f_s/2$ tones extracted from I and Q tributaries. Meanwhile, $g_{\text{Rx,}f_s/4}^2$ can be calculated by the power ratio between the extracted two $f_s/4$ tones from I and Q tributaries. It can improve the accuracy by filtering out the noise at the frequencies outside $f_s/2$ and $f_s/4$. Finally, the more accurate Rx IQ power imbalance $g^2_{\text{Rx,IQ}}$ can be obtained by averaging the $g_{\text{Rx,}f_s/2}^2$ and $g_{\text{Rx,}f_s/4}^2$.

\begin{figure*}[t]
\centering
\includegraphics[width= \linewidth]{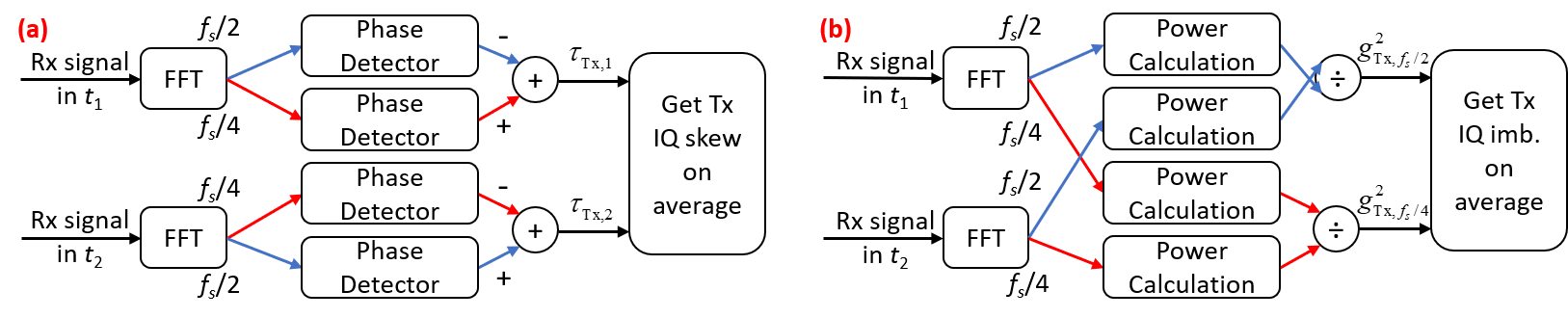}
\caption{Schematic diagram of far-end IQ-impairments estimation based on the specially designed TFITs including  Tx IQ (a) skew and (b) power imbalance.}
\label{fig_5}
\end{figure*}

\subsection{Estimation of Tx IQ Impairments in PtMP Networks}
After compensating for the Rx IQ impairments, the Rx TFITs suffer from the Tx IQ impairments, which can be expressed as Eqs. (\ref{eq12})-(\ref{eq14}). $s_{\text{Rx}}(t)$ in the time slot $t_i$ is denoted as $s_{\text{Rx},{t_i}}(t)$ where $i = 1, 2$, which can be expressed as
\begin{equation}\label{eq28}
s_{\text{Rx},{t_1}}(t)= s_{\text{Rx},f_s/2,t_1}(t)+js_{\text{Rx},f_s/4,t_1}(t)
\end{equation}
and
\begin{equation}\label{eq29}
s_{\text{Rx},{t_2}}(t)= s_{\text{Rx},f_s/4,t_2}(t)+js_{\text{Rx},f_s/2,t_2}(t),
\end{equation}
respectively. In the time slot $t_1$, the $s_{\text{Rx},f_s/4,t_1}(t)$ suffers from the Tx IQ skew and power imbalance. In the time slot $t_2$, the $s_{\text{Rx},f_s/2,t_2}(t)$ suffers from the Tx IQ skew and power imbalance. Based on the model of IQ impairments, we propose TFITs-based schemes for accurately estimating the Tx IQ skew and power imbalance.

\subsubsection{Estimation of Tx IQ Skew}
In the time slot $t_1$, $s_{\text{Rx},f_s/2, t_1}(t)$ in the Eq. (\ref{eq13}) suffers from the timing phase offset caused by the out-of-synchronization clock, while $s_{\text{Rx},f_s/4, t_1}(t)$ in the Eq. (\ref{eq14}) suffers from both the timing phase offset of the clock and Tx IQ skew $\tau_{\text{Tx}}$. Therefore, the Tx IQ skew $\tau_{\text{Tx}}$ can be calculated by the difference in timing phase offset between $f_s/4$ and $f_s/2$ tones in the time slot $t_1$. Similarly, the Tx IQ skew can be also calculated by the difference in timing phase offset between $f_s/2$ and $f_s/4$ tones in the time slot $t_2$.

Fig. \ref{fig_5}(a) shows the schematic diagram of Tx IQ skew estimation based on the TFITs. By performing sampling and FFT on the Rx signal $s_{\text{Rx},t_1}(t)$ in the time slots $t_1$ to $S_{\text{Rx},t_1}(k)$ and $s_{\text{Rx},t_2}(t)$ in the time slots $t_2$ to $S_{\text{Rx},t_2}(k)$, the timing phase offset of $f_s/2$ and $f_s/4$ tones are estimated by the Godard phase detector and the modified Godard phase detector, respectively. Thus, the Tx IQ skew can be estimated by subtracting the timing phase offset of the $f_s/2$ and $f_s/4$ tones in the time slot $t_1$ as 
\begin{equation}\label{eq30}
\begin{aligned}
\tau_{\text{Tx},t_{1}} & =  \tau_{\text{Tx}} + \Delta \tau(f_s/4,~f_s/2) \\ &  = \frac{1}{\pi} \operatorname{arg}\left[\sum_{k=3 N / 8-n+1}^{3 N / 8+n} S_{\text{Rx},t_1}(k) S_{\text{Rx},t_1}^*\left(k+\frac{N}{4}\right)\right] \\
& -\frac{1}{2 \pi} \operatorname{arg}\left[\sum_{k=N / 4-n+1}^{N / 4+n} S_{\text{Rx},t_1}(k) S_{\text{Rx},t_1}^*\left(k+\frac{N}{2}\right)\right]
\end{aligned}
\end{equation}
where $\Delta \tau(f_s/4,~f_s/2)$ denotes the additional timing offset caused by phase frequency response between the $f_s/2$ and $f_s/4$ tones. The Tx IQ skew can be also estimated by using the tones at $f_s/2$ and $f_s/4$ at time slot $t_2$ as 
\begin{equation}\label{eq31}
\begin{aligned}
\tau_{\text{Tx},t_{2}} & =  \tau_{\text{Tx}} + \Delta \tau(f_s/2,~f_s/4) \\ & =\frac{1}{2 \pi} \operatorname{arg}\left[\sum_{k=N / 4-n+1}^{N / 4+n} S_{\text{Rx},t_2}(k) S_{\text{Rx},t_2}^*\left(k+\frac{N}{2}\right)\right] \\
& -\frac{1}{\pi} \operatorname{arg}\left[\sum_{k=3N / 8-n+1}^{3N / 8+n} S_{\text{Rx},t_2}(k) S_{\text{Rx},t_2}^*\left(k+\frac{N}{4}\right)\right]
\end{aligned}
\end{equation}
where $\Delta \tau(f_s/2,~f_s/4)$ denotes the additional timing offset caused by phase frequency response between $f_s/2$ and $f_s/4$ tones. Since the frequencies of tones are interleaved at the two time slots, $\Delta \tau(f_s/4,~f_s/2)$ is equal to $-\Delta \tau(f_s/2,~f_s/4)$. Therefore, the Tx IQ skew $\tau_{\text{Tx,IQ}}$ can be accurately acquired by averaging $\tau_{\text{Tx},t_{1}}$ and $\tau_{\text{Tx},t_{2}}$ at time slots $t_1$ and $t_2$. 

\subsubsection{Estimation of Tx IQ Power Imbalance}
The Tx IQ power imbalance can be calculated by the power ratio between two same-frequency tones at two time slots. The power ratio between two $fs/2$ tones at the $t_2$ and $t_1$ time slots can be calculated by
\begin{equation}\label{eq32}
g_{\text{Tx,}f_s/2}^2=\frac{E\left(\left| s_{\text{Rx},{f_s/2,t_2}}(t)\right|^2\right)}{E\left(\left|s_{\text{Rx},{f_s/2,t_1}}(t)\right|^2\right)} =g_{\text{Tx}}^2\frac{P_{f_s/2,t_2}}{P_{f_s/2,t_1}}
\end{equation}
where $P_{f_s/2,t_1}$ and $P_{f_s/2,t_2}$ denote the average power of the $f_s/2$ tones without power imbalance in the time slots $t_1$ and $t_2$, respectively. Similarly, the power ratio between two $f_s/4$ tones at the $t_1$ and $t_2$ time slots can be calculated by
\begin{equation}\label{eq33}
g_{\text{Tx,}f_s/4}^2=\frac{E\left(\left| s_{\text{Rx},{f_s/4,t_1}}(t)\right|^2\right)}{E\left(\left|s_{\text{Rx},{f_s/4,t_2}}(t)\right|^2\right)}=g_{\text{Tx}}^2\frac{P_{f_s/4,t_1}}{P_{f_s/4,t_2}}
\end{equation}
where $P_{f_s/4,t_1}$ and $P_{f_s/4,t_2}$ denote the average power of the $f_s/4$ tones without power imbalance in the time slots $t_1$ and $t_2$, respectively. The divisions in Eqs. (\ref{eq32}) and (\ref{eq33}) can be calculated by
\begin{equation}\label{eq34}
\frac{P_{f_s/2,t_2}}{P_{f_s/2,t_1}} = \frac{{E\left(|I_{f_s/2}(t+\tau_{\text{Tx}})|^2+|Q_{f_s/2}(t+\tau_{\text{Tx}})|^2\right)}}{{E\left(|I_{f_s/2}(t)|^2+|Q_{f_s/2}(t)|^2\right)}}=1
\end{equation}
and
\begin{equation}\label{eq35}
\frac{P_{f_s/4,t_1}}{P_{f_s/4,t_2}} = \frac{{E\left(|I_{f_s/4}(t+\tau_{\text{Tx}})|^2+|Q_{f_s/4}(t+\tau_{\text{Tx}})|^2\right)}}{{E\left(|I_{f_s/4}(t)|^2+|Q_{f_s/4}(t)|^2\right)}}=1,
\end{equation}
respectively. Fig. \ref{fig_5}(b) shows the estimation of Tx IQ power imbalance based on the TFITs. After the FFT, the average powers of $f_s/2$ and $f_s/4$ tones at the two time slots are separately calculated. It can improve the accuracy by filtering out the noise at the frequencies outside of $f_s/2$ and $f_s/4$. The Tx IQ power imbalance $g_{\text{Tx,IQ}}^2$ can be obtained by $g_{\text{Tx,}f_s/2}^2$ and $g_{\text{Tx,}f_s/4}^2$. Finally, the estimation accuracy of $g_{\text{Tx,IQ}}^2$ can be improved by averaging the $g_{\text{Tx,}f_s/2}^2$ and $g_{\text{Tx,}f_s/4}^2$.

\begin{figure*}[t]
\centering
\includegraphics[width=\linewidth]{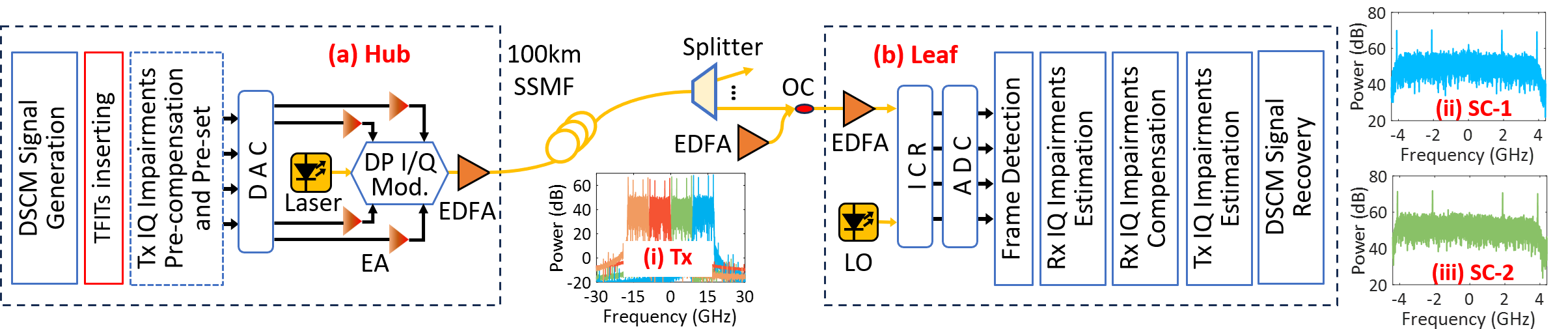}
\caption{The experimental setups of the coherent optical PtMP network to verify the feasibility of the TFITs-based far-end IQ-impairments estimation. Inset (i) is the electrical spectrum of the DSCM signal at the hub. Insets (ii) and (iii) are the electrical spectra of the received SC-1 and SC-2 at the leaf.}
\label{fig_6}
\end{figure*}

\section{Experimental Setups}
Fig. \ref{fig_6} shows experimental setups of an 8Gbaud/SC$\times$4SCs DSCM-based coherent PtMP optical network to verify the feasibility of the TFITs-based far-end IQ-impairments estimation. The bit sequence was mapped to 16 quadrature amplitude modulation (16QAM) symbols at the hub. Then, each subcarrier used a root-raised cosine (RRC) pulse-shaping filter with a roll-off factor of 0.1. After the frequency upshift, the four subcarriers were multiplexed to generate a DSCM signal. The specially designed TFITs were inserted before the DSCM signal for estimating the IQ impairments at the expense of spectral efficiency. Owing to the slowly changing IQ impairments, we can transmit the TFITs at intervals such as every online registration, not every data frame. In our experiment, one block of TFITs in Fig. \ref{fig_3} contained 2048 symbols. We averaged the IQ impairments estimated by 3 blocks of TFITs to achieve higher estimation accuracy. A pre-compensation of Tx IQ impairments for the DSCM signal was implemented at the hub based on the far-end estimated IQ impairments at the leaf. The Tx IQ power imbalance can be compensated by multiplying an estimated coefficient $1/g_{\text{Tx,IQ}}$ on the Q tributary. Meanwhile, the Tx IQ skew can be effectively compensated by interpolation based on $\tau_{\text{Tx,IQ}}$.

\begin{figure*}[t]
\centering
\includegraphics[width=\linewidth]{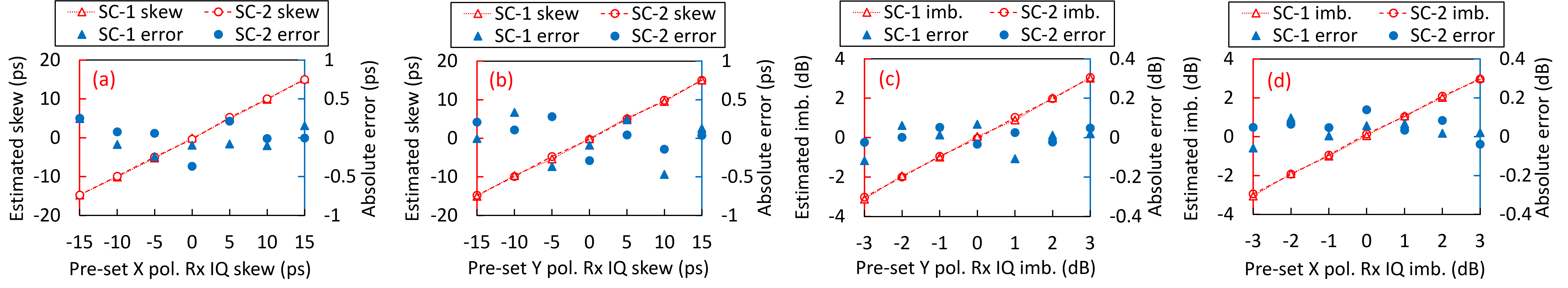}
\caption{Estimated Rx IQ impairments versus pre-set Rx IQ impairments and the absolute error between the estimated Rx IQ impairments and pre-set Rx IQ impairments. (a) X polarization Rx IQ skew, (b) Y polarization Rx IQ skew, (c) X polarization Rx IQ power imbalance, and (d) Y polarization Rx IQ imbalance. The Pol. denotes Polarization. The Imb. denotes Imbalance.}
\label{fig_7}
\end{figure*}

To verify the universality of the TFITs-based far-end IQ-impairments estimation, pre-set Tx IQ impairments were added to the digital DSCM signal. After adding pre-set Tx IQ impairments, a digital-to-analog converter (DAC) converted the digital signal to an analog signal with a sampling rate of 90GSa/s. Inset (i) in Fig. \ref{fig_6} shows the electrical spectrum of the DSCM signal at the hub. An external cavity laser (ECL) operating at $1550.12$nm with a less than 100kHz linewidth was used as the optical carrier. After being amplified by electrical amplifiers (EAs), the analog signal drove a dual-polarizations IQ modulator (DP I/Q Mod.) to generate the optical signal. The output optical power was approximately $-14$dBm. After being boosted by an Erbium-doped fiber amplifier (EDFA) to approximately 6 dBm, the optical signal was launched into a 100 km standard single-mode fiber (SSMF).

The optical signal is coupled with the optical white noise generated by EDFA by a 90:10 optical coupler (OC) to adjust the optical SNR (OSNR). At the leaf, the received optical signal was amplified by an EDFA to achieve an optical power of $-10$dBm. Then, the optical signal was detected by an integrated coherent receiver (ICR) to obtain the electrical signals. A tunable ECL with a linewidth less than 100 kHz was used as the LO, which has an output optical power of approximately $12$dBm. The LO wavelength was tuned to match the central wavelength of the subcarrier for dropping the corresponding subband \cite{zhang2012towards}. The electrical signals were digitized by a 90GSa/s analog-to-digital converter (ADC) and then recovered by the Rx digital signal processing (DSP). 

Insets (ii) and (iii) in Fig. \ref{fig_6} are the received spectra of the SC-1 and SC-2 at the leaf, respectively. Without loss of generality, SC-1 and SC-2 were used to estimate the IQ impairments. After frame detection, Rx IQ skew and power imbalance were estimated based on the specially designed TFITs and proposed algorithms. The Rx IQ power imbalance can be compensated by multiplying an estimated coefficient $1/g_{\text{Rx,IQ}}$ in Eq. (\ref{eq21}) on the Q tributary. Meanwhile, the Rx IQ skew can be effectively compensated by interpolation based on $\tau_{\text{Rx,IQ}}$. After compensating Rx IQ impairments, the Tx IQ skew $\tau_{\text{Tx,IQ}}$ and power imbalance $g_{\text{Tx,IQ}}$ were estimated by the specially designed TFITs and proposed algorithms and then transmitted to the hub using the uplink. The estimated Tx IQ skews and power imbalances can be averaged at the hub to achieve a more accurate IQ skew and power imbalance. After the IQ impairments estimation, the DSCM signal was recovered by DSP as demonstrated in Refs. \cite{wang2023fast,10403903, wang2024non}.

\begin{figure*}[t]
\centering
\includegraphics[width=\linewidth]{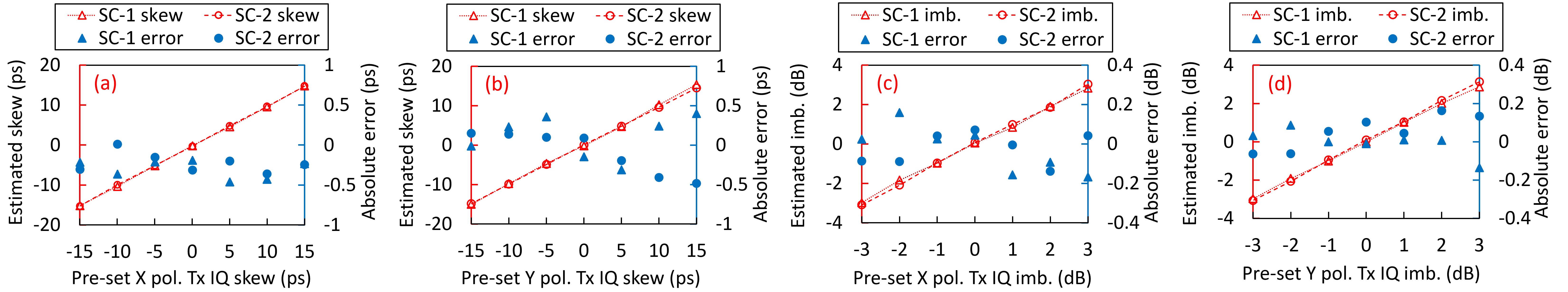}
\caption{Estimated Tx IQ impairments versus pre-set Tx IQ impairments and the absolute error between the estimated Tx IQ impairments and pre-set Tx IQ impairments. (a) X polarization Tx IQ skew, (b) Y polarization Tx IQ skew, (c) X polarization Tx IQ imbalance, and (d) Y polarization Tx IQ imbalance.}
\label{fig_8}
\end{figure*}
 
\section{Experimental Results and Discussions}
Firstly, the TFITs-based IQ-impairments estimation is evaluated when only IQ skew or power imbalance exists. The OSNR was set to approximately 17dB for verifying the robustness of TFITs-based IQ-impairments estimation under large enough noise. Figs. \ref{fig_7}(a) and (b) show the estimated Rx IQ skews versus pre-set Rx IQ skews and the absolute errors between the estimated Rx IQ skews and pre-set Rx IQ skews for X and Y polarizations, respectively. When the Rx IQ skews were set to the values from $-15$ps to $15$ps (i.e. 24\% of the symbol duration) on X and Y polarizations, the TFITs-based Rx IQ skew estimation can achieve the absolute error within $\pm0.5$ps. Figs. \ref{fig_7}(c) and (d) show the estimated Rx IQ power imbalances versus pre-set Rx IQ power imbalances and the absolute errors between the estimated Rx IQ power imbalances and pre-set Rx IQ power imbalances for X and Y polarizations, respectively. When the Rx IQ imbalances were set to the values from $-3$dB to $3$dB, the TFITs-based Rx IQ power imbalance estimation can achieve an absolute error within $\pm0.2$dB. 

Figs. \ref{fig_8}(a) and (b) show the estimated Tx IQ skews versus pre-set Tx IQ skews and the absolute errors between the estimated Tx IQ skews and pre-set Tx IQ skews for the X and Y polarizations, respectively. When only the Tx IQ skews were set to the values from $-15$ps to $15$ps on X and Y polarizations, the TFITs-based far-end Tx IQ-skew estimation can achieve the absolute error within $\pm0.5$ps. Figs. \ref{fig_8}(c) and (d) depict the estimated Tx IQ power imbalances versus pre-set Tx IQ power imbalances and the absolute errors between the estimated Tx IQ power imbalances and pre-set Tx IQ power imbalances for the X and Y polarizations, respectively. When the Tx IQ power imbalances were set to the values from $-3$dB to $3$dB, the TFITs-based far-end Tx IQ-power-imbalance estimation can achieve an absolute error within $\pm0.2$dB. 

\begin{figure*}[t]
\centering
\includegraphics[width=\linewidth]{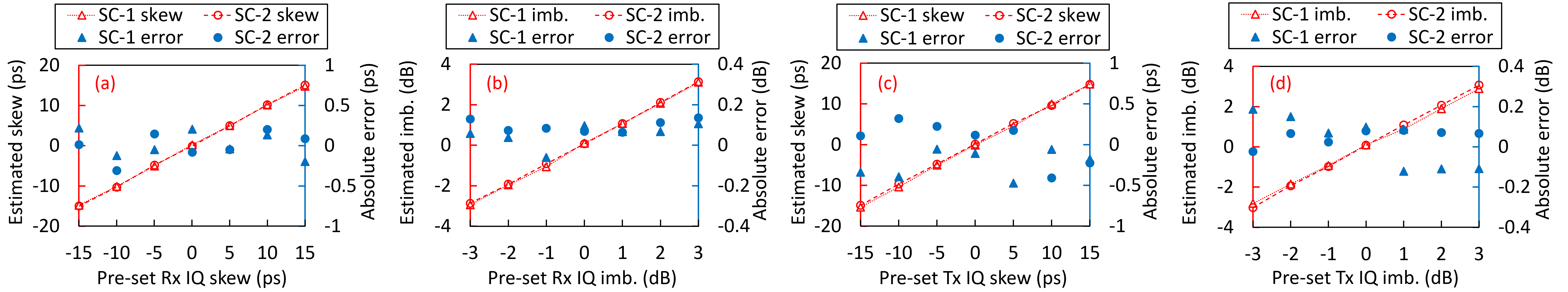}
\caption{Estimated IQ impairments versus pre-set IQ impairments and the absolute error between the estimated IQ impairments and pre-set IQ impairments under (a) $5$ps Tx IQ skew, $1$dB Tx IQ power imbalance, and $-1$dB Rx IQ power imbalance, (b) $5$ps Tx IQ skew, $1$dB Tx IQ power imbalance, and $-5$ps Rx IQ skew, (c) $1$dB Tx IQ power imbalance, $-5$ps Rx IQ skew, and $-1$dB Rx IQ power imbalance, and (d) $5$ps Tx IQ skew, $-5$ps Rx IQ skew, and $-1$dB Rx IQ power imbalance, respectively.}
\label{fig_9}
\end{figure*}

The TFITs-based IQ-impairments estimation is also evaluated when several IQ impairments coexist. Fig. \ref{fig_9}(a) shows the estimated Rx IQ skew versus pre-set Rx IQ skew, and the absolute error between the estimated Rx IQ skew and pre-set Rx IQ skew when $5$ps Tx IQ skew, $1$dB Tx IQ power imbalance, and $-1$dB Rx IQ power imbalance coexist. The Rx IQ skew from $-15$ps to $15$ps can be estimated with the absolute error within $\pm0.5$ps by using TFITs-based Rx IQ-skew estimation. Fig. \ref{fig_9}(b) shows the estimated Rx IQ power imbalance versus pre-set Rx IQ power imbalance, and the absolute error between estimated Rx IQ power imbalance and pre-set Rx IQ power imbalance when $5$ps Tx IQ skew, $1$dB Tx IQ power imbalance, and $-5$ps Rx IQ skew coexist. The Rx IQ power imbalance from $-3$dB to $3$dB can be estimated with the absolute error within $\pm0.2$dB by using TFITs-based Rx IQ-power-imbalance estimation.

Fig. \ref{fig_9}(c) shows the estimated Tx IQ skew versus pre-set Tx IQ skew and the absolute error between the estimated Tx IQ skew and pre-set Tx IQ skew when $1$dB Tx IQ power imbalance, $-5$ps Rx IQ skew, and $-1$dB Rx IQ power imbalance coexist. The Tx IQ skew from $-15$ps to $15$ps can be estimated with the absolute error within $\pm0.5$ps by TFITs-based far-end Tx IQ-skew estimation. Fig. \ref{fig_9}(d) shows the estimated Tx IQ power imbalance versus pre-set IQ power imbalance and the absolute error between estimated IQ power imbalance and pre-set IQ power imbalance when $5$ps Tx IQ skew, $-5$ps Rx IQ skew, and $-1$dB Rx IQ power imbalance coexist. The Tx IQ power imbalance from $-3$dB to $3$dB can be estimated with the absolute error within $\pm0.2$dB by TFITs-based far-end Tx IQ-power-imbalance estimation. In conclusion, TFITs-based IQ-impairments estimation can accurately estimate Rx and Tx IQ impairments at the far-end leaf.

\begin{figure*}[t]
\centering
\includegraphics[width=\linewidth]{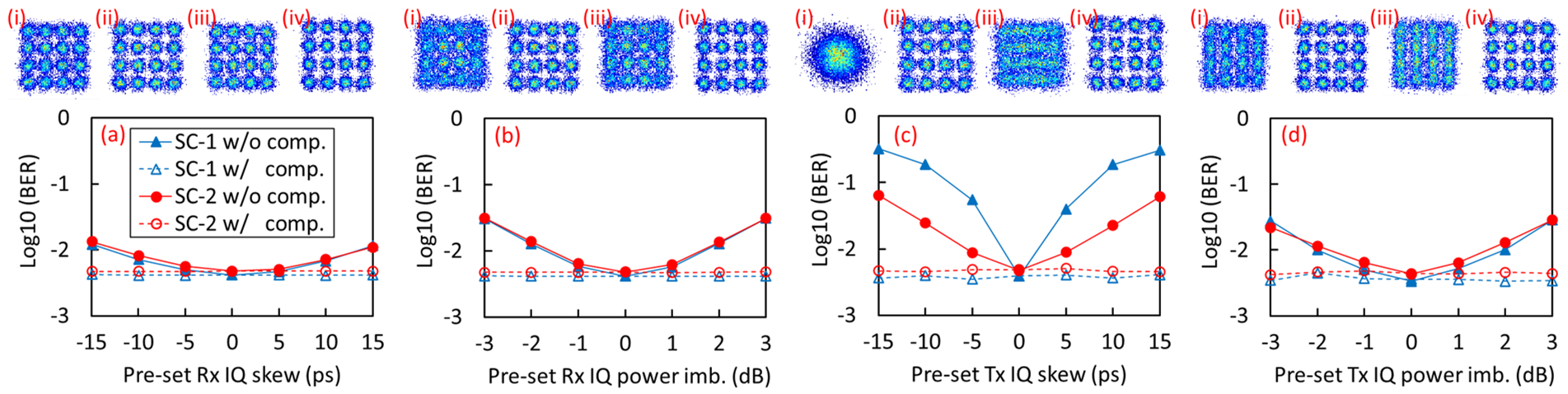}
\caption{BER performance at SC-1 and SC-2 versus (a) Rx IQ skew, (b) Rx IQ power imbalance, (c) Tx IQ skew, (d) IQ power imbalance without (w/o) and with (w/) compensation (comp.). Insets are the constellation diagrams of (i) SC-1 without compensation, (ii) SC-1 with compensation, (iii) SC-2 without compensation, and (iv) SC-2 with compensation, respectively.}
\label{fig_10}
\end{figure*}

In $8$Gbaud/SC×$4$SCs DSCM-based coherent PtMP optical network, the bit-error ratios (BERs) of SC-1 and SC-2 are used to evaluate the effect of IQ-impairments estimation and compensation. The OSNR was set to approximately 22dB to ensure the BER can achieve $0.0038$. As Figs. \ref{fig_10}(a) and (b) show, the BERs of the SC-1 and SC-2 increase with the increase of Rx IQ skew or power imbalance. Insets of Figs. \ref{fig_10}(a) and (b) depict the constellation diagrams under $15$ps Rx IQ skew or $3$dB Rx IQ power imbalance of (i) SC-1 without compensation, (ii) SC-1 with compensation, (iii) SC-2 without compensation, and (iv) SC-2 with compensation, respectively. After compensation for Rx IQ impairments, the BER penalties caused by Rx IQ skews or power imbalances are almost eliminated, and the constellations are clearer. Significantly, the impacts of the Rx IQ impairments on the BERs of the SC-1 and SC-2 are similar because the two subcarriers are detected as the baseband signals at the leaf.

Figs. \ref{fig_10}(c) and (d) depict the BERs of the SC-1 and SC-2 under the Tx IQ skews and Tx IQ power imbalances, respectively. Insets of Figs. \ref{fig_10}(c) and (d) show the constellation diagrams under $15$ps Tx IQ skew or $3$dB Tx IQ power imbalance of (i) SC-1 without compensation, (ii) SC-1 with compensation, (iii) SC-2 without compensation, and (iv) SC-2 with compensation, respectively. Since the image crosstalk caused by Tx IQ skew increases with the frequency increase, higher-frequency SC-1 suffers from more distortions than lower-frequency SC-2. Thus, the BER performance of SC-1 is more sensitive to Tx IQ skew than that of SC-2. Since the image crosstalk caused by the Tx IQ power imbalance is the same on all the frequencies, the higher-frequency SC-1 suffers from almost the same distortions as the lower-frequency SC-2, leading to the same trend of BER performance degradation. After compensation for Tx IQ impairments, the BER penalty on the two subcarriers can be almost eliminated, and the constellations become much clearer. In conclusion, the proposed IQ-impairments estimation and compensation can effectively improve the BER performance for the DSCM-based coherent PtMP optical networks.

\section{Conclusion}
In this paper, the TFITs are designed to implement the far-end IQ-impairments estimation for DSCM-based coherent PtMP optical networks. The TFITs-based far-end IQ-impairments estimation does not require the signals on two symmetrical subcarriers, which can estimate Tx IQ skew and power imbalance of hub at an individual leaf. The $8$Gbaud/SC×$4$SCs DSCM-based coherent PtMP optical network is experimentally demonstrated to verify the feasibility of TFITs-based IQ-impairments estimation. The experimental results depict that TFITs-based IQ-impairments estimation can achieve absolute errors in IQ skew and power imbalance within $\pm0.5$ps and $\pm0.2$dB, respectively. After compensation for the IQ impairments, the BER performance can be significantly improved, especially the high-frequency subcarriers. In conclusion, TFITs-based IQ-impairments estimation has great potential for DSCM-based coherent PtMP optical networks.

It is worth noting that the current version of our proposed estimation cannot address the challenge of frequency-dependent IQ skew and power imbalance. We propose a possible improved scheme based on more TFITs to address this challenge. The number of TFITs on one subcarrier can be increased to scale down the gaps between the adjacent frequency tones. When the gaps are small enough, the TFITs suffer almost the same IQ skew and power imbalance, which can be used to estimate the IQ skew and power imbalance on the corresponding frequency points. Meanwhile, the IQ skew and power imbalance should be compensated individually for the corresponding frequency points. In this way, longer lengths of TFITs and larger sizes of FFT should be employed to achieve high estimation accuracy.

\appendices
\section{Impact of Phase Quadrature Error on IQ Skew and Power Imbalance Estimation}
This appendix will prove the impact of phase quadrature error on the IQ skew and imbalance estimation. Firstly, We investigate the model of IQ impairments including IQ skew, power imbalance, and phase quadrature error on the TFITs in the time slot $t_1$. The TFITs with Tx IQ impairments can be modeled as\cite{fludger2016transmitter}
\begin{equation}\label{eq36}
s_{\text{Tx},\theta_{\text{Tx}}}(t)=s_{\text{Tx},{f_s/2}}(t)+je^{j\theta_{\text{Tx}}}g_{\text{Tx}}s_{\text{Tx},{f_s/4}}(t+\tau_{\text{Tx}})
\end{equation}
where $\tau_{\text{Tx}}$, $g^2_{\text{Tx}}$, and $\theta_\text{Tx}$ represent the Tx IQ skew, power imbalance, and phase quadrature error, respectively. $s_{\text{Tx},{f_s/2}}(t)$ and $s_{\text{Tx},{f_s/4}}(t)$ are defined as Eq. (\ref{eq8}). The I and Q tributaries of $s_{\text{Tx},\theta_{\text{Tx}}}(t)$ can be expressed as
\begin{equation}\label{eq38}
s_{\text{Tx},\theta_{\text{Tx}},\text{I}}(t)=s_{\text{Tx},{fs/2}}(t)-g_{\text{Tx}}\sin\theta_{\text{Tx}}s_{\text{Tx},{f_s/4}}(t+\tau_{\text{Tx}})
\end{equation}
and
\begin{equation}\label{eq39}
s_{\text{Tx},\theta_{\text{Tx}},\text{Q}}(t)=g_{\text{Tx}}\cos\theta_{\text{Tx}}s_{\text{Tx},{f_s/4}}(t+\tau_{\text{Tx}}),
\end{equation}
respectively. After optical modulation, the launch optical signal can be defined as
\begin{equation}\label{eq37}
s_{\text{Tx},\theta_{\text{Tx}},\text{O}}(t)=\left[s_{\text{Tx},\theta_{\text{Tx}},\text{I}}(t)+js_{\text{Tx},\theta_{\text{Tx}},\text{Q}}(t)\right]e^{j2\pi f_0t}
\end{equation}
where $f_0$ is the center frequency of the optical carrier.

At the Rx side, we select SC-1 for the frequency downshift. The received signal of SC-1 with Tx IQ impairments can be expressed as 
\begin{equation}\label{eq40}
\begin{aligned}
s_{\text{Rx},{\theta_{\text{Tx}}}}(t)=&\left[s_{\text{Tx},\theta_{\text{Tx}},\text{I}}(t)+js_{\text{Tx},\theta_{\text{Tx}},\text{Q}}(t)\right]e^{-j \left[ 2 \pi (f_\text{c} +\Delta f)t+\phi\right]}\\=& s_{\text{Rx},{f_s/2}}(t)+je^{j\theta_{\text{Tx}}}s_{\text{Rx},{f_s/4}}(t)
\end{aligned}
\end{equation}
where $s_{\text{Rx},{f_s/2}}(t)$ and $s_{\text{Rx},{f_s/4}}(t)$ are defined as Eqs. (\ref{eq13}) and (\ref{eq14}). Both the I and Q tributaries of the received signal $s_{\text{Rx},{\theta_{\text{Tx}}}}(t)$ contain the $f_s/4$ and $f_s/2$ tones. The I tributary of $s_{\text{Rx},{\theta_{\text{Tx}}}}(t)$ can be calculated by
\begin{equation}\label{eq41}
r_{f_s/2,\theta_{\text{Tx}},\text{I}}(t)=r_{f_s/2,\text{I}}(t)
\end{equation}
plus
\begin{equation}\label{eq42}
r_{f_s/4,\theta_{\text{Tx}},\text{I}}(t)=\cos\theta_{\text{Tx}}r_{f_s/4,\text{I}}(t)-\sin\theta_{\text{Tx}}r_{f_s/4,\text{Q}}(t).
\end{equation}
The Q tributary of $s_{\text{Rx},{\theta_{\text{Tx}}}}(t)$ can be calculated by
\begin{equation}\label{eq43}
r_{f_s/2,\theta_{\text{Tx}},\text{Q}}(t)=r_{f_s/2,\text{Q}}(t)
\end{equation}
plus
\begin{equation}\label{eq44}
r_{f_s/4,\theta_{\text{Tx}},\text{Q}}(t)=\sin\theta_{\text{Tx}}r_{f_s/4,\text{I}}(t)+\cos\theta_{\text{Tx}}r_{f_s/4,\text{Q}}(t).
\end{equation}
where $r_{f_s/2, \text{I}}(t)$, $r_{f_s/4, \text{I}}(t)$, $r_{f_s/2, \text{Q}}(t)$, and $r_{f_s/4, \text{Q}}(t)$ are defined as Eqs. (\ref{eq13}) and (\ref{eq14}). Therefore, the I and Q tributaries of the signal $s_{\text{Rx},\theta_{\text{Tx}}}(t)$ can be expressed as
\begin{equation}\label{eq45}
s_{\text{Rx},{\theta_{\text{Tx}}},\text{I}}(t)= r_{f_s/2,\theta_{\text{Tx}},\text{I}}(t)+r_{f_s/4,\theta_{\text{Tx}},\text{I}}(t)
\end{equation}
and
\begin{equation}\label{eq46}
s_{\text{Rx},\theta_{\text{Tx}},\text{Q}}(t)=r_{f_s/2,\theta_{\text{Tx}},\text{Q}}(t)+r_{f_s/4,\theta_{\text{Tx}},\text{Q}}(t),
\end{equation}
respectively. When the received signal suffers the Rx IQ impairments, it can be modeled as\cite{fludger2016transmitter}
\begin{equation}\label{eq47}
s_{\text{Rx},\theta_{\text{Rx}}}(t)= s_{\text{Rx},\theta_{\text{Rx}},\text{I}}(t) + js_{\text{Rx},\theta_{\text{Rx}},\text{Q}}(t) 
\end{equation}
where $s_{\text{Rx},\theta_{\text{Rx}},\text{I}}(t)$ and $s_{\text{Rx},\theta_{\text{Rx}},\text{Q}}(t)$ are defined as 
\begin{equation}\label{eq48}
s_{\text{Rx},\theta_{\text{Rx}},\text{I}}(t) = s_{\text{Rx},{\theta_{\text{Tx}}},\text{I}}(t)
\end{equation}
and
\begin{equation}\label{eq49}
s_{\text{Rx},\theta_{\text{Rx}},\text{Q}}(t)=g_{\text{Rx}}\begin{bmatrix} \cos\theta_{\text{Rx}} & -\sin\theta_{\text{Rx}} \end{bmatrix} \begin{bmatrix} s_{\text{Rx},\theta_{\text{Tx}},\text{Q}}(t+\tau_{\text{Rx}}) \\ s_{\text{Rx},\theta_{\text{Tx}},\text{I}}(t+\tau_{\text{Rx}}) \end{bmatrix},
\end{equation}
respectively where $\tau_{\text{Rx}}$, $g_{\text{Rx}}^2$, and $\theta_{\text{Rx}}$ denote the Rx IQ skew, power imbalance and phase quadrature error, respectively.

\subsubsection{Estimation of Rx IQ Skew}
By substituting Eqs. (\ref{eq45}) and (\ref{eq46}) into Eqs. (\ref{eq48}) and (\ref{eq49}), it is inferred that in the time slot $t_1$, $r_{f_s/2,\theta_{\text{Tx}},\text{I}}(t)$ contains only the timing phase offset caused by the out-of-synchronization clock, while $-g_{\text{Rx}}\sin\theta_{\text{Rx}}r_{f_s/2,\theta_{\text{Tx}}\text{I}}(t+\tau_{\text{Rx}})+g_{\text{Rx}}\cos\theta_{\text{Rx}}r_{f_s/2,\theta_{\text{Tx}},\text{Q}}(t+\tau_{\text{Rx}})$ contains both the timing phase offset of the clocks and Rx IQ skew $\tau_{\text{Rx}}$. the difference between the timing phase offsets of $r_{f_s/2,\theta_{\text{Tx}},\text{I}}(t)$ and $-g_{\text{Rx}}\sin\theta_{\text{Rx}}r_{f_s/2,\theta_{\text{Tx}},\text{I}}(t+\tau_{\text{Rx}})+g_{\text{Rx}}\cos\theta_{\text{Rx}}r_{f_s/2,\theta_{\text{Tx}},\text{Q}}(t+\tau_{\text{Rx}})$ is the  Rx IQ skew $\tau_{\text{Rx}}$. Similarly, the difference between the timing phase offsets of $r_{f_s/4,\theta_{\text{Tx}},\text{I}}(t)$ and $-g_{\text{Rx}}\sin\theta_{\text{Rx}}r_{f_s/4,\theta_{\text{Tx}},\text{I}}(t+\tau_{\text{Rx}})+g_{\text{Rx}}\cos\theta_{\text{Rx}}r_{f_s/4,\theta_{\text{Tx}},\text{Q}}(t+\tau_{\text{Rx}})$ is also the Rx IQ skew $\tau_{\text{Rx}}$. Therefore, the Rx IQ skew can be estimated by calculating the difference in the timing phase offset between the Q and the I tributaries of two same-frequency tones, which indicates that the proposed Rx IQ skew estimation is not influenced by phase quadrature error.
\subsubsection{Estimation of Rx IQ Power Imbalance}
According to Eqs. (\ref{eq48}) and (\ref{eq49}), the power ratio between the Q and I tributaries of $f_s/2$ tones can be calculated by
\begin{equation}\label{eq50}
\begin{aligned}
&g_{\text{Rx,}f_s/2}^2\\&=\frac{E\left[\left| g_{\text{Rx}}\cos\theta_{\text{Rx}}r_{f_s/2,\theta_{\text{Tx}},\text{Q}}(t)-g_{\text{Rx}}\sin\theta_{\text{Rx}}r_{f_s/2,\theta_{\text{Tx}},\text{I}}(t)\right|^2\right]}{E\left[\left|r_{f_s/2, \theta_{\text{Tx}},\text{I}}(t)\right|^2\right]}\\&
=g_{\text{Rx}}^2 \frac{E\left[ {\sin^2\theta_{\text{Rx}}}\left|r_{f_s/2,\text{I}}(t)\right|^2+\cos^2\theta_{\text{Rx}}\left|r_{f_s/2,\text{Q}}(t)\right|^2\right]}{E\left[\left|r_{f_s/2, \text{I}}(t)\right|^2\right]}
\end{aligned}
\end{equation}
Since the existing frequency offset causes a varied $\phi_1(t)$ and $\phi_2(t)$ with a rotation of $2\pi$, $E\left[\left|r_{f_s/2,\text{I}}(t)\right|^2\right]$ is equal to $E\left[\left|r_{f_s/2,\text{Q}}(t)\right|^2\right]$ according to Eq. (\ref{eq13}). Therefore, $g_{\text{Rx,}f_s/2}^2$ can be used to estimate the adding Rx IQ power imbalance $g_{\text{Rx}}^2$. According to Eqs. (\ref{eq48}) and (\ref{eq49}), the power ratio between the Q and I tributaries of $f_s/4$ tones can be calculated by 
\begin{equation}\label{eq51}
\begin{aligned}
&g_{\text{Rx,}f_s/4}^2\\&=
g_{\text{Rx}}^2 \frac{E\left[ {\sin^2\Delta\theta}\left|r_{f_s/4,\text{I}}(t)\right|^2+\cos^2\Delta\theta\left|r_{f_s/4,\text{Q}}(t)\right|^2\right]}{E\left[ {\cos^2\theta_{\text{Tx}}}\left|r_{f_s/4,\text{I}}(t)\right|^2+\sin^2\theta_{\text{Tx}}\left|r_{f_s/4,\text{Q}}(t)\right|^2\right]}
\end{aligned}
\end{equation}
where $\Delta\theta = \theta_{\text{Rx}}-\theta_{\text{Tx}}$. Since the existing frequency offset causes a varied $\phi_1(t)$ and $\phi_2(t)$ with a rotation of $2\pi$, $E\left[\left|r_{f_s/4,\text{I}}(t)\right|^2\right]$ is equal to $E\left[\left|r_{f_s/4,\text{Q}}(t)\right|^2\right]$ according to Eq. (\ref{eq14}). Therefore, $g_{\text{Rx,}f_s/4}^2$ can be used to estimate the adding Rx IQ power imbalance $g_{\text{Rx}}^2$. In conclusion, the Rx IQ power imbalance can be accurately estimated under the phase quadrature error.

\subsubsection{Estimation of Tx IQ Skew}
The well-known Gram-Schmidt orthogonalization procedure (GSOP)\cite{fatadin2008compensation} can compensate for Rx phase quadrature error. After compensation for the Rx IQ impairments, the Rx IFITs suffer from the Tx impairments, which can be expressed as Eq. (\ref{eq40}). According to Eq. (\ref{eq36}), it is inferred that in the time slot $t_1$, $f_s/2$ tone suffers from the timing phase offset caused by the out-of-synchronization clock, while $f_s/4$ tone suffers from both the timing phase offset of the clock and Tx IQ skew $\tau_{\text{Tx}}$. The Tx IQ skew $\tau_{\text{Tx}}$ can be calculated by the difference in timing phase offset between $f_s/4$ and $f_s/2$ tones in the time slot $t_1$. Similarly, the same analysis can be performed for the tones in the time slot $t_2$. Therefore, phase quadrature error does not influence the Tx IQ skew estimation.

\subsubsection{Estimation of Tx IQ Power Imbalance}
$s_{\text{Rx},\theta_{\text{Tx}}}(t)$ in the time slot $t_i$ is denoted as $s_{\text{Rx},\theta_{\text{Tx}},{t_i}}(t)$ where $i = 1, 2$, which can be expressed as
\begin{equation}\label{eq52}
s_{\text{Rx},\theta_{\text{Tx}},{t_1}}(t)= s_{\text{Rx},{f_s/2},{t_1}}(t)+je^{j\theta_{\text{Tx}}}s_{\text{Rx},{f_s/4},{t_1}}(t)
\end{equation}
and
\begin{equation}\label{eq53}
s_{\text{Rx},\theta_{\text{Tx}},{t_2}}(t)= s_{\text{Rx},{f_s/4},{t_2}}(t)+je^{j\theta_{\text{Tx}}}s_{\text{Rx},{f_s/2},{t_2}}(t),
\end{equation}
respectively. The power ratio between two $f_s/2$ tones at the $t_2$ and $t_1$ time slots can be calculated by
\begin{equation}\label{eq54}
g_{\text{Tx,}f_s/2}^2=\frac{E\left(\left|e^{j\theta_{\text{Tx}}}s_{\text{Rx},{f_s/2,t_2}}(t)\right|^2\right)}{E\left(\left|s_{\text{Rx},{f_s/2,t_1}}(t)\right|^2\right)} =g_{\text{Tx}}^2
\end{equation}
Similarly, the same analysis can be performed for the $f_s/4$ tones. Therefore, the Tx IQ power imbalance can be calculated by the power ratio between two same-frequency tones at two time slots, which indicates that the Rx IQ power imbalance can be accurately estimated under the phase quadrature error.

\bibliographystyle{ieeetr}
\bibliography{Ref}
\end{document}